\documentstyle[preprint,aps]{revtex}
\begin{document}
\draft
\title
{ Negative dimensional operators in the disordered critical points of
Dirac fermions }
\author{Claudio de C. Chamon, Christopher Mudry and Xiao-Gang Wen}
\address{Department of Physics, Massachusetts Institute of Technology,
77 Massachusetts Avenue, Cambridge, Massachusetts 02139}
\date{December 23, 1994}
\maketitle

\begin{abstract}
Recently, in an attempt to study disordered criticality
in Quantum Hall systems and $d$-wave superconductivity,
it was found that two dimensional random Dirac fermion systems
contain a line of critical points which is connected to
the pure system. We use bosonization and current algebra to study
properties of the critical line and calculate the exact scaling dimensions
of all local operators. We find that the critical line
contains an infinite number of relevant operators with negative
scaling dimensions.

\end{abstract}
\narrowtext

The transitions between quantum Hall (QH) states induced by disorder
have been a long standing problem \cite{QH-transition}.
Recently, it was pointed out that the transitions
can also be induced by periodic potentials.
The critical properties of these transitions in pure systems
can then be studied through
$1/N$ expansion and perturbative expansion
\cite{Wen 1993,CFW}.
In particular, in absence of interactions,
the pure critical point at the transition (between integer QH (IQH) states)
can be described by
an effective theory for Dirac fermions in 1+2 space-time dimensions.
It is then natural to investigate what happens
to the pure critical point if disorder is included.
This point of view was taken recently by Ludwig et al.
\cite{Ludwig 1994}.

In this letter we plan to study some interesting effects of impurities
on the pure critical point of IQH transition.
Our starting point is the Dirac Hamiltonian in two spatial dimensions
for the low energy spectrum of the {\it pure system at criticality}:
$$
H^{\ }_0=
-i\ \gamma^{\ }_{\mu}\partial^{\ }_{\mu}.
$$
Here, $\mu=1,2$ denote the two spatial directions and
$
(\gamma^{\ }_1,\gamma^{\ }_2,\gamma^{\ }_5)=
(\sigma^{\ }_1,\sigma^{\ }_2,\sigma^{\ }_3)
$
are the Pauli matrices. Three types of disorder are important:
$$
H^{\ }_{\rm imp}=
-
\sqrt{g^{\ }_A} A^{\ }_{\mu}(x) \gamma^{\ }_{\mu}
+
\sqrt{g^{\ }_V} V(x)
+
\sqrt{g^{\ }_M} M(x) \gamma^{\ }_5.
$$
They correspond to random gauge potential, random chemical potential,
and random mass. (The constant mass $\sqrt{g^{\ }_M}M_0$ is the
parameter that controls the transition in the pure system \cite{Ludwig
1994}.)  The strength of the impurities is parameterized by three
{\it positive} parameters $g^{\ }_{A,V,M}$, respectively, assuming the
fluctuations of $A_\mu$, $V(x)$, and $M(x)$ to be given by $
exp-{1\over2}\int d^2 x (A^2_{\mu} +V^2+M^2)$. To leading order in
the impurities strength, the three randomness represent marginal
perturbations to the pure critical point $H^{\ }_0$.  Including the
second order terms and in presence of only one type of impurities, it
was shown in Ref. \cite{Ludwig 1994} that $g^{\ }_M$ is marginally
irrelevant, $g^{\ }_V$ is marginally relevant, and $g^{\ }_A$ is exactly
marginal to all orders and generate a line of critical points starting
at the free Dirac fermion model. It is the properties of this critical
line that we are going to concentrate on in this letter.  The main result
obtained here is that all the critical points on the critical line
(except for the free Dirac fermion model with $g^{\ }_A=0$) contain
infinitely many operators with {\it negative} dimensions. Those
operators can be generated as the higher order terms in a non-Gaussian
distribution of the random gauge potentials.  Thus the critical points
on the critical line are unstable in a very special sense -- {\it they
have infinitely many relevant directions}.

The same Hamiltonian $H=H^{\ }_0+H^{\ }_{\rm imp}$
and its generalizations obtained
by allowing for internal symmetries such as spin or isospin
describe many non-interacting two-dimensional electronic systems
characterized by isolated Fermi points
in the presence of static disorder.
Examples are degenerate semi-conductors,
two-dimensional graphite sheets,
tight-binding Hamiltonians in the flux phase,
and dirty $d$-wave superconductors in two-dimensions
\cite{Examples,Nersesyan 1994}.


Some exact properties of the critical line have been studied
through bosonization in the replica approach\cite{Ludwig 1994,Nersesyan 1994}.
In this letter we will
use the supersymmetric formalism for non-interacting disordered
systems \cite{Efetov 1983}.
The advantage of the supersymmetric approach
is that the operator content of the critical points can be
easily obtained. This allows us to study
a large class of {\it local} operators that may appear in the low energy
effective theory.

In the supersymmetric formalism, one begins by representing the
{\it unaveraged} one-particle Green function at frequency $\omega$
by a path integral over pairs of Grassmann spinors
$\bar\psi,\psi$ and pairs of complex spinors
$\bar\varphi,\varphi$:
\begin{eqnarray}
G^{\pm}_{\omega}(x,y)&=&
\langle
x\left|
{1\over\omega-H\pm i 0^+}
\right| y
\rangle
\label{unaveragedpropagator}
\\
&=&
\pm i
\int\!{\cal D}[\bar\psi   ,\psi   ]
\int\!{\cal D}[\bar\varphi,\varphi]
\psi(y)\bar\psi(x)\ e^{i\int d^2x{\cal L}^{\pm}_\omega},
\nonumber
\end{eqnarray}
where
$$
{\cal L}^{\pm}_\omega =
\bar\psi
\left[
(\pm)(\omega-H)+i0^+
\right]
\psi
+
\bar\varphi
\left[
(\pm)(\omega-H)+i0^+
\right]
\varphi.
$$
Integration over
$\bar\varphi$ and $\varphi$ gives the denominator
in the usual path-integral representation of Green functions.
The infinitesimal positive number $0^+$ insures the
convergence of the bosonic path integral.
The impurity averaged (retarded) Green function is then given by
the RHS of Eq. (\ref{unaveragedpropagator})
with ${\cal L}^{\pm}_{\omega}$ replaced by
\begin{eqnarray}
{\cal L}^{\rm eff}_{\omega}&=&
\bar\psi
\left[
i\gamma^{\ }_{\mu}\partial^{\ }_{\mu}
+
\omega
+i0^+
\right]
\psi
+
\bar\varphi
\left[
i\gamma^{\ }_{\mu}\partial^{\ }_{\mu}
+
\omega
+i0^+
\right]
\varphi
\nonumber\\
&+&\
i{g^{\ }_A\over2}
\left(
\bar\psi\gamma^{\ }_{\mu}\psi
+
\bar\varphi\gamma^{\ }_{\mu}\varphi
\right)^2
+
i{g^{\ }_V\over2}
\left(
\bar\psi\psi
+
\bar\varphi\varphi
\right)^2
\nonumber\\
&+&\
i{g^{\ }_M\over2}
\left(
\bar\psi\gamma^{\ }_5
\psi
+
\bar\varphi\gamma^{\ }_5
\varphi
\right)^2.
\label{effectivesupersymmetricaction}
\nonumber
\end{eqnarray}
Impurity averaging has turned a non-interacting problem into
an interacting one.
The effective action ${\cal L}^{\rm eff}_{\omega}$ has an internal $U(1/1)$
graded symmetry (supersymmetry).

The effective theory can be solved exactly if
$\omega=g^{\ }_1=g^{\ }_M=g^{\ }_V=0$.
In this limit, the effective action defines a
supersymmetric generalization of the Thirring model that we solve
exactly in two ways.  One approach is based on the current algebra
with $U(1/1)$ graded internal symmetry. It is quite general.  We apply
it to the particular case of the $U(1/1)$ Thirring model.  Another
approach is specifically suited to the Thirring model, where it is
possible to decouple the vector impurity from the spinors.  In both
cases the operator content on the critical line is calculated.  This
will allow us to assess the stability of the line of critical points.
We begin with the latter approach.

The partition function for the $U(1/1)$ Thirring model is equivalent to
averaging the one-particle Green function over impurities coupling to
the currents only
\begin{eqnarray}
Z^{\rm Th}&=&
\int\!{\cal D}[\bar\psi   ,\psi   ]\!
\int\!{\cal D}[\bar\varphi,\varphi]\!
\int\!{\cal D}[\bar\alpha,\alpha]\
e^{-\bar\alpha\partial^{2 }\alpha}\!
\int{ {\cal D}[\Phi^{\ }_1,\Phi^{\ }_2]\over{\cal M}}\!
\nonumber\\
&\times&
\exp
\Big\{
-
\bar\psi
\gamma^{\ }_{\mu}
\left[
\partial^{\ }_{\mu}-
{\rm i}\sqrt{g^{\ }_A}
\left(
\tilde\partial^{\ }_{\mu}\Phi^{\ }_1+\partial^{\ }_{\mu}\Phi^{\ }_2
\right)
\right]
\psi
\nonumber\\
&-&
\bar\varphi
\gamma^{\ }_{\mu}
\left[
\partial^{\ }_{\mu}-
{\rm i}\sqrt{g^{\ }_A}
\left(
\tilde\partial^{\ }_{\mu}\Phi^{\ }_1+\partial^{\ }_{\mu}\Phi^{\ }_2
\right)
\right]
\varphi
\nonumber\\
&-&
{1\over2}
\left[
\left(\partial^{\ }_{\mu}\Phi^{\ }_1\right)^2
+
\left(\partial^{\ }_{\mu}\Phi^{\ }_2\right)^2
\right]
\Big\}.
\nonumber
\end{eqnarray}
Here, we have rewritten the random gauge potential $A^{\ }_{\mu}$ as
$
A^{\ }_{\mu}=
\tilde\partial^{\ }_{\mu}\Phi^{\ }_1+
      \partial^{\ }_{\mu}\Phi^{\ }_2
$
($\tilde\partial^{\ }_{\mu}=\epsilon^{\ }_{\mu\nu}\partial^{\ }_{\mu}$,
$\epsilon^{\ }_{12}=-\epsilon_{21}=1$).
We have also
reexponentiated the Jacobian ${\rm Det}\ \partial^2$ for this change of
variables with the help of two ghost fields $\bar\alpha$ and $\alpha$
so as to preserve the unity of the partition function (${\cal M}$ normalizes
the measure of the impurity).
One can decouple the spinors from the impurity potentials through a
redefinition of the spinor fields.  In terms of the chiral components
$\psi_\pm=\frac{1}{2}(1\pm\gamma_5)\psi$, the decoupling
transformation is:
\begin{eqnarray}
\psi ^{\dag}_{\pm}&=&
\psi'^{\dag}_{\pm}
e^{\mp\sqrt{g^{\ }_A}\Phi^{\ }_1- {\rm i}\sqrt{g^{\ }_A}\Phi^{\ }_2},\;
\psi ^{\   }_{\pm}= 
e^{\pm\sqrt{g^{\ }_A}\Phi^{\ }_1+ {\rm i}\sqrt{g^{\ }_A}\Phi^{\ }_2}
\psi'_{\pm},
\nonumber\\
\varphi ^{\dag}_{\pm}&=&
\varphi'^{\dag}_{\pm}
e^{\mp\sqrt{g^{\ }_A}\Phi^{\ }_1- {\rm i}\sqrt{g^{\ }_A}\Phi^{\ }_2},\;
\varphi^{\   }_{\pm}= 
e^{\pm\sqrt{g^{\ }_A}\Phi^{\ }_1+ {\rm i}\sqrt{g^{\ }_A}\Phi^{\ }_2}
\varphi'^{\   }_{\pm}.
\nonumber
\end{eqnarray}
This transformation leaves the measure unchanged due to the supersymmetry.
We thus map the (interacting)
$U(1/1)$ Thirring model onto three independent and free sectors
\begin{eqnarray}
Z^{\rm Th}&=&
\int
{\cal D}[\psi'^{\dag}_{\pm},\psi'_{\pm}]
{\cal D}[\varphi'^{\dag}_{\pm},\varphi'_{\pm}]
\nonumber\\
&\times&
e^
{
-2\left(
\psi'^{\dag}_+
\partial^{\ }_{\bar z}
\psi'_+
+
\psi'^{\dag}_-
\partial^{\ }_z
\psi'_-
+
\varphi'^{\dag}_+
\partial^{\ }_{\bar z}
\varphi'_+
+
\varphi'^{\dag}_-
\partial^{\ }_z
\varphi'_-
\right)
}
\nonumber\\
&\times&\
\int{\cal D}[\bar\alpha,\alpha]
e^
{
-
\bar\alpha\ \partial^2\ \alpha
}
\nonumber\\
&\times&
\int{ {\cal D}[\Phi^{\ }_1,\Phi^{\ }_2]\over{\cal M} }\
e^
{
-
{1\over2}
\left[
\left(\partial^{\ }_{\mu}\Phi^{\ }_1\right)^2
+
\left(\partial^{\ }_{\mu}\Phi^{\ }_2\right)^2
\right]
},
\nonumber
\end{eqnarray}
where we use complex coordinates $z=x_1+ix_2$ and ${\bar z}=x_1-ix_2$.
Each sector is {\it conformally invariant}.
The impurity strength does not appear explicitly, because it is hidden
in the rotation from the original fields to the new ones.

Any local operator in the original variables can be rewritten in terms of
the four primary fields $\psi'^{\   }_+,\cdots,\varphi'^{\ }_-$ and the two
primary fields
$e^{i(i\sqrt{g^{\ }_A})\Phi^{\ }_1}$
and
$e^{i  \sqrt{g^{\ }_A} \Phi^{\ }_2}$.
Whereas the scaling dimension of
$e^{i  \sqrt{g^{\ }_A} \Phi^{\ }_2}$
is positive,
the scaling dimension of
$e^{i(i\sqrt{g^{\ }_A})\Phi^{\ }_1}$
is negative
due to the non-compactness of the decoupling
transformation. This in itself is not surprising.
For example, primary fields with negative dimensions are common place
in the treatment of strongly correlated electronic system \cite{Mudry 1994}.
They can turn a marginal interaction such as the Umklapp term
into a marginally relevant interaction as the current-current interaction
increases beyond a critical value \cite{Haldane 1982}.
However, in the context of our supersymmetric model it leads
for any given value of $g^{\ }_A>0$ to the existence
of infinitely many local operators which are compatible with the $U(1/1)$
symmetry and which carry negative dimensions. This is in sharp contrast
to unitary theories (say the U(1) Thirring model) where
operators with negative dimensions are not allowed.

To see this, we begin by calculating the two point functions.
The only non-vanishing ones are
\begin{eqnarray}
\langle \psi_+(z) \psi^{\dag}_+(0) \rangle &=&
\frac{1}{2\pi z}=
\langle \varphi_+(z) \varphi^{\dag}_+(0) \rangle\ ,
\nonumber
\\
\langle \psi_-({\bar z}) \psi^{\dag}_-(0) \rangle &=&
\frac{1}{2\pi{\bar z}}=
\langle \varphi_-({\bar z}) \varphi^{\dag}_-(0) \rangle\ ,
\label{propagator}
\end{eqnarray}
as follows immediately from, say,
\begin{eqnarray}
\langle
\psi^{\dag}_{\pm}(x)
\psi^{\   }_{\pm}(0)
\rangle
&=&
\langle
\psi'^{\dag}_{\pm}(x)
\psi'^{\   }_{\pm}(0)
\rangle
\nonumber\\
&\times&
\langle
e^{ i (- i \sqrt{g^{\ }_A})\Phi^{\ }_1(x)}
e^{ i (+ i \sqrt{g^{\ }_A})\Phi^{\ }_1(0)}
\rangle
\nonumber\\
&\times&
\langle
e^{ i (-       \sqrt{g^{\ }_A})\Phi^{\ }_2(x)}
e^{ i (+       \sqrt{g^{\ }_A})\Phi^{\ }_2(0)}
\rangle .
\label{producttwopoint}
\end{eqnarray}
At the level of the two-point function,
the scaling properties of the $\psi$ and the $\varphi$ are not changed by
the randomness of the gauge potential.
However, this cannot not be the case for two-point functions of
composite operators. For example, consider the composite operator ${\cal O}=$
$
\psi_+^{\dag m^{\ }_1}
\psi_-^{\dag m^{\ }_2}
\times
$
$\cdots$
$
\times
\varphi_+^{m^{\ }_7}
\varphi_-^{m^{\ }_8},
$
which is defined through point splitting.
We find, with a calculation along the lines of  Eq. (\ref{producttwopoint}),
that ${\cal O}$ scales like
\cite{Knizhnik 1984,Mudry 1994}
\begin{equation}
\langle
{\cal O}^{\dag}(z,{\bar z})
{\cal O}^{\   }(0)
\rangle
\ \propto\
       z^{-2h}
{\bar z}^{-2{\bar h}}
\label{scalingofvertexoperator},
\end{equation}
where the two conformal weights $(h,{\bar h})$ are given by
\begin{eqnarray}
&&
h=
{1\over2}
\left[
m^2_1+m^2_3+m^{\ }_5+m^{\ }_7
+
\left(
f^2_2
-
|f^{\ }_1|^2
\right)
{g^{\ }_A\over4\pi}
\right],
\nonumber
\\
&&
{\bar h}=
{1\over2}
\left[
m^2_2+m^2_4+m^{\ }_6+m^{\ }_8
+
\left(
f^2_2
-
|f^{\ }_1|^2
\right)
{g^{\ }_A\over4\pi}
\right].
\nonumber
\end{eqnarray}
Here $f_{1,2}$ are purely imaginary and real integer functions,
respectively:
\begin{eqnarray}
&&
f_1=
- i
\left(
m^{\ }_1-m^{\ }_2-m^{\ }_3+m^{\ }_4
+
m^{\ }_5-m^{\ }_6-m^{\ }_7+m^{\ }_8
\right),
\nonumber
\\
&&
f_2=
-\
\left(
m^{\ }_1+m^{\ }_2-m^{\ }_3-m^{\ }_4
+
m^{\ }_5+m^{\ }_6-m^{\ }_7-m^{\ }_8
\right).
\nonumber
\end{eqnarray}

Negative dimensional operators are now easily obtained.
We consider the composite operator $\Psi^{\ }_{n^{\ }_1n^{\ }_2}$ defined by
\begin{equation}
\Psi^{\ }_{n^{\ }_1n^{\ }_2}=
\cases{
\varphi^{n^{\ }_1}_+ \varphi^{n^{\ }_2}_-,&$n^{\ }_1>0,n^{\ }_2>0$,\cr
\varphi_+^{\dag-n_1} \varphi^{n^{\ }_2}_-,&$n^{\ }_1<0,n^{\ }_2>0$,\cr
\varphi^{n_1}_+      \varphi_-^{\dag-n_2},&$n^{\ }_1>0,n^{\ }_2<0$,\cr
\varphi_+^{\dag-n_1} \varphi_-^{\dag-n_2},&$n^{\ }_1<0,n^{\ }_2<0$,\cr}
\label{Psi}
\end{equation}
and apply the result above.
The conformal weights of these
operators are
\begin{equation}
      h =\frac{1}{2}|n^{\ }_1|+\frac{g^{\ }_A}{2\pi}n^{\ }_1n^{\ }_2,\;
{\bar h}=\frac{1}{2}|n^{\ }_2|+\frac{g^{\ }_A}{2\pi}n^{\ }_1n^{\ }_2.
\label{weightsofPsi}
\end{equation}
This demonstrates that for {\it any given} value of $g^{\ }_A$,
there are infinitely many local composite operators with
negative conformal weights. This is very different from the $U(1)$
Thirring model which is a unitary theory and thus cannot support operators
with negative dimensions.
Furthermore, operators like  $\Psi^{\dag}_{-|n||n|}$ are generated in
the effective supersymmetric model if non-Gaussian moments in the
probability distribution of the $M(x)$ or $V(x)$ are present.

We now formulate a current algebra description of the model Eq.
(\ref{effectivesupersymmetricaction}) along the critical line
$\omega=g^{\ }_M=g^{\ }_V=0$.  This provides us with an independent check of
Eq.
(\ref{weightsofPsi}), and more importantly, can be generalized to any
conformally invariant theory with internal $U(1/1)$ supersymmetry.  We first
construct the current algebra description for the free model, and then
show that the current algebra approach also applies to any
positive value of $g^{\ }_A$.

When $g^{\ }_A=0$ we have the free theory
$$
{\cal L}^{\ }_0= 2i\left(
\psi^{\dag}_+\partial_{\bar z}\psi^{\ }_+ +
\psi^{\dag}_-\partial_z\psi^{\ }_- +
\varphi^{\dag}_+\partial_{\bar z}\varphi^{\ }_+ +
\varphi^{\dag}_-\partial_z\varphi^{\ }_-
\right).
$$
It describes two decoupled sectors labeled by the $\pm$ subscripts of
the chiral components.
One can study this theory by canonically quantizing it along
contours of equal ``time'' in the $x^{\ }_1$-$x^{\ }_2$ Euclidean plane.
In what follows, we shall use, without loss of generality, equal $x_2$
lines. The canonical momenta associated with the fields $\psi_\pm$ are
$$
\Pi^{\ }_{\psi_\pm}=
\frac{\partial {\cal L}_0}{\partial (\partial_2\psi_\pm)}=
\mp \psi_\pm^{\dag},
\ \
\Pi^{\ }_{\varphi_\pm}=
\frac{\partial {\cal L}_0}{\partial (\partial_2\varphi_\pm)}=
\mp \varphi_\pm^{\dag},
$$
and the equal $x^{\ }_2$ (``time'') quantization conditions are
$\{\psi_\pm(x_1,x_2) ,
\Pi_{\psi_\pm}(x'_1,x_2)\}=i\delta(x_1-x'_1)$
and $[\varphi_\pm(x_1,x_2) ,
\Pi_{\varphi_\pm}(x'_1,x_2)]=i\delta(x_1-x'_1)$. The time-ordered
Green's functions for $\psi,\varphi$ can be obtained from the
classical equations of motion found from the Lagrangian, together with
the above commutation relations used as boundary conditions. One
recovers the two-point functions of Eqs. (\ref{propagator})
obtained from the path integral.

The free model has an internal graded symmetry $U(1/1)$, generated by two
bosonic and two fermionic currents. The bosonic $U(1)$ symmetries are
$
(\psi^{\ }_\pm,\varphi^{\ }_\pm)\rightarrow
(e^{i\theta^{\ }_\psi   }\psi   ^{\ }_\pm
,e^{i\theta^{\ }_\varphi}\varphi^{\ }_\pm)
$.
The fermionic symmetries are generated by the infinitesimal
transformations
$
(\delta\psi^{\ }_\pm,\delta\varphi^{\ }_\pm)=
(\theta\varphi^{\ }_\pm,-{\bar \theta}\psi^{\ }_\pm)
$,
where $\theta$ and ${\bar \theta}$ are Grassmann variables. The normal
ordered conserved currents associated with these symmetries are:
\begin{eqnarray}
J^\psi_\pm(x_1,x_2)&\equiv&
\psi^{\dag}_\pm(x_1+\epsilon,x_2)\psi_\pm(x_1-\epsilon,x_2)-
\frac{1}{4\pi\epsilon},
\nonumber\\
J^\varphi_\pm(x_1,x_2)&\equiv&
\varphi^{\dag}_\pm(x_1+\epsilon,x_2)\varphi_\pm(x_1-\epsilon,x_2)+
\frac{1}{4\pi\epsilon},\nonumber
\\
\eta_\pm(x_1,x_2)&\equiv&
\varphi^{\dag}_\pm(x_1+\epsilon,x_2)\psi_\pm(x_1-\epsilon,x_2),
\nonumber\\
{\bar \eta}_\pm(x_1,x_2)&\equiv&
\psi^{\dag}_\pm(x_1+\epsilon,x_2)\varphi_\pm(x_1-\epsilon,x_2),
\nonumber
\end{eqnarray}
where point splitting has been used ($\epsilon\rightarrow 0$). It is
convenient to define the currents $J_\pm=J_\pm^\psi+J_\pm^\varphi$ and
$j_\pm=J_\pm^\psi-J_\pm^\varphi$.  Using the equal $x_2$ commutation
relations for the $\psi$,$\varphi$'s, we find that the non-vanishing
commutators are:
\begin{eqnarray}
[J_\pm(x_1,x_2),j_\pm(x'_1,x_2)]&=&\pm \frac{i}{\pi}\delta'(x_1-x'_1),\nonumber
\cr
[j_\pm(x_1,x_2),\eta_\pm(x'_1,x_2)]&=&
\pm 2 i\ \delta(x_1-x'_1)\ \eta_\pm(x'_1,x_2), \cr
[j_\pm(x_1,x_2),{\bar \eta}_\pm(x'_1,x_2)]&=&
\mp 2 i\ \delta(x_1-x'_1)\ {\bar \eta}_\pm(x'_1,x_2),
\nonumber \\
\{\eta_\pm(x_1,x_2),{\bar \eta}_\pm(x'_1,x_2)\}&=&
\mp i
\Big[
\frac{1}{2\pi}\delta'(x_1-x'_1)\nonumber \\
&+&
\delta(x_1-x'_1)\ J_\pm(x'_1,x_2)
\Big].
\nonumber
\end{eqnarray}
All commutators between the $+$ and $-$ sectors vanish.  The chiral
components $J^{\ }_+$ and $J^{\ }_-$ are the holomorphic and antiholomorphic
components of $J_1=J_+ +J_-$ and $J_{2}=i(J_+ - J_-)$, respectively.

{}From the equations of motion derived using ${\cal L}^{\ }_0$,
we find that fields in the $+$ ($-$) sectors
such as
$\psi^{\ }_{\pm}$,
$\varphi^{\ }_{\pm}$,
and
$I^i_{\pm}=J^{\ }_{\pm},j^{\ }_{\pm},\eta^{\ }_{\pm},{\bar \eta}^{\ }_{\pm}$,
depend only on $z$ ($\bar z$). Such fields are called chiral fields.
The same result can also be obtained from the quantum equation of motion
$\partial_2{\cal O}=i[H^0,{\cal O}]$, where $H^0$ is derived from our
canonical quantization. However, it is more convenient to express
$H^0$ as a bilinear form in the currents. We find that $H^0=H^0_++H^0_-$ with
\begin{equation}
H^0_\pm=i\pi\int dx_1\ \left(J_\pm J_\pm + J_\pm j_\pm
+ \eta_\pm{\bar \eta}_\pm - {\bar \eta}_\pm\eta_\pm\right)
\end{equation}
exactly reproduces the equations of motion for the currents and
$\psi$, $\varphi$.
It then follows that, for chiral fields, say in the $+$ sector,
$
{\cal O}^{\ }_+(z)=
$
$
e^{+H^0_+z}{\cal O}^{\ }_+(0)e^{-H^0_+z}
$.
Notice that $H^0$ is invariant under $U(1/1)$ transformations.

The impurity averaging induces an interaction
${\cal L_{\rm int}}=i\frac{g^{\ }_A}{2}J^2_\mu=i2g^{\ }_AJ_+J_-$
in the Lagrangian. This interaction,
in turn, corresponds to
${\cal H}_{\rm int}=-i2g^{\ }_A\ J_+J_-$.
The new Hamiltonian can be diagonalized, {\it while}
keeping the $+$ and $-$ sectors independent, by a redefinition of
the currents
${\widetilde J}_\pm=J_\pm$,
${\widetilde \eta}_\pm=\eta_\pm$,
${\bar {\widetilde \eta}}_\pm={\bar \eta}_\pm$,
${\widetilde j}_\pm=j_\pm-\frac{g^{\ }_A}{\pi}J_\pm -\frac{g^{\
}_A}{\pi}J_\mp$.
For convenience, we have chosen to maintain the $j$ charge density
unchanged, {\it i.e.}, ${\widetilde j}_2=i({\widetilde
j}_+-{\widetilde j}_-)=i(j_+-j_-)=j_2$. The chiral components
$H^{\ }_{\pm}$ of $H=\int dx_1 {\cal H}$ are now given by
\begin{equation}
{{\cal H}^{\ }_\pm\over i\pi}=
\left(1+\frac{g^{\ }_A}{\pi}\right)
{\widetilde J}_\pm{\widetilde J}_\pm              +
{\widetilde J}_\pm{\widetilde j}_\pm              +
{\widetilde \eta}_\pm{\bar {\widetilde \eta}}_\pm -
{\bar {\widetilde \eta}}_\pm{\widetilde \eta}_\pm .
\label{rotatedH}
\end{equation}
With this redefinitions of the currents, one can check that the only
current commutator that changes is $[{\widetilde j}_\pm(x_1,x_2),
{\widetilde j}_\pm(x'_1,x_2)]=\mp i\frac{2g^{\ }_A}{\pi^2}\
\delta'(x_1-x'_1)$. We thus see that the interacting theory also
decouples into a holomorphic and antiholomorphic sector.
The Hamiltonian Eq. (\ref{rotatedH})
and the commutators between $\tilde I^i_{\pm}$ also
allow us to calculate the exact $N$-point correlations between the
currents. One can show that all the correlations have algebraic decay,
which implies that the interacting model describes a critical point.
This demonstrates the conformal invariance of the theory for arbitrary
values of $g^{\ }_A$.

We now sketch how the current algebra approach can be used to
calculate the anomalous dimension of a generic operator ${\cal O}$
constructed locally from powers of $\psi$'s and $\varphi$.
Due to the conformal symmetry of the
critical point,
the two-point correlation
function on the plane, Eq. (\ref{scalingofvertexoperator}),
can be mapped onto the one on a
cylinder: $\langle {\cal O}(z,\bar z) {\cal O}(0)\rangle \sim
e^{ihz-i{\bar h}\bar z}$ for large $x_2$. We have taken
the circular direction of the
cylinder to be $0\leq x_1 <2\pi$.
{}From
\begin{equation}
\langle {\cal O}(z,\bar z) {\cal O}(0)\rangle =
\langle 0|{\cal O}(0)e^{-zH_+ +\bar z H_-} {\cal O}(0)|0\rangle,
\label{Heisenberg2-point}
\end{equation}
we see that the large $x_2$ behavior is controlled by the
state $|{\cal O}\rangle$. Here, we have expanded ${\cal O}(0)|0\rangle$ into
the eigenstates of $iH$ [$(iH)^\dagger=iH$], and $|{\cal O}\rangle$
is the one with the minimal eigenvalue
$E^{{\cal O}^{\ }_+}_0+E^{{\cal O}^{\ }_-}_0$,
where $H_\pm |{\cal O}\rangle = E^{{\cal O}^{\ }_\pm}_0 |{\cal O}\rangle$.
The dimensions of ${\cal O}$ are
then simply given by $(h,{\bar h})=(E^{{\cal O}_+}_0,E^{{\cal
O}_-}_0)$.

Notice now that the Fourier components
of the currents, ${\tilde I}^j_{\pm;n}=i\int dx_1\ {\tilde
I}^j_\pm(x_1)e^{inx_1}$, satisfy $[iH,{\tilde I}^j_{\pm;n}]=\pm
n{\tilde I}^j_{\pm;n}$. Hence, the modes ${\tilde I}^j_{\pm;n}$ raise
and lower the eigenvalues of $iH$.
One can then show that the state $|{\cal O}\rangle$
is annihilated by the lowering operators:
${\tilde I}^j_{+;-n}|{\cal O}\rangle ={\tilde I}^j_{-;n}|{\cal
O}\rangle=0$, for $n>0$,
so that using Eq. (\ref{rotatedH}),
\begin{eqnarray}
iH_\pm|{\cal O}\rangle=
\frac{1}{2}
\Big[\
(1&+&\frac{g^{\ }_A}{\pi}){\widetilde J}_{\pm;0}{\widetilde J}_{\pm;0} +
{\widetilde J}_{\pm;0} {\widetilde j}_{\pm;0}\nonumber\\
&+&{\widetilde \eta}_{\pm;0}{\bar {\widetilde \eta}}_{\pm;0} -
{\bar {\widetilde \eta}}_{\pm;0}{\widetilde \eta}_{\pm;0}\Big]
|{\cal O}\rangle .\label{HWweights}
\end{eqnarray}
By reexpressing Eq. (\ref{HWweights}) in terms of the unrotated currents,
one finally obtains
\begin{equation}
h |{\cal O}\rangle=
h_{\rm free}|{\cal O}\rangle-\frac{g^{\ }_A}{2\pi}J_{+;0}J_{-;0}|{\cal
O}\rangle,
\end{equation}
where $h_{\rm free}$ is the weight of ${\cal O}$ in the absence of
interactions. A similar expression can be found for ${\bar h}$. Thus
we reach the very simple result that the conformal weights $(h,\bar
h)$ of ${\cal O}$ can be calculated from the charges of ${\cal O}$ for
unrotated currents.

The $J$ charges of operators constructed from the $\psi,\varphi$'s can
be easily obtained from their commutators with $J_{\pm;0}$. The
conformal weights for $\Psi^{\ }_{n^{\ }_1n^{\ }_2}$ in Eq.
(\ref{weightsofPsi}) then follows immediately by recognizing that its
unrotated charges are $(J_{+;0},J_{-;0})=(-n_1,n_2)$ and that its free
conformal weights are $(h_{\rm free},{\bar h}_{\rm
free})=(\frac{1}{2}|n_1|,\frac{1}{2}|n_2|)$.


We believe that negative dimensional operators are generic to
disordered critical points, as one can infer from this work and the
$2+\epsilon$ expansion \cite{Wegner 1980}.
In general, negative dimensional operators are allowed
since supersymmetric models for the disordered
critical points are not unitary and, as such,
contain states with negative norm.
Therefore, in the search for physical disordered critical points,
it is crucial to find under what conditions
the generally allowed negative dimensional
operators do not spoil the critical points.

We would also like to point out that the supersymmetric model studied
here can be viewed as the fermionized disordered $X\!-\!Y$ model studied in
\cite{Rubinstein 1982}.
Our results are then very suggestive of a similar instability that
might occur along the critical line of the disordered $X\!-\!Y$ model
\cite{Korshunov 1993}. A more detailed study will appear elsewhere.

We would like to thank B. Altshuler, M. Kardar, P. A. Lee, and Z. Q. Wang
for sharing their insights.
This work was supported by NSF grants  DMR-9411574 (XGW) and
DMR-9400334 (CCC).
CM acknowledges a fellowship from the Swiss Nationalfonds and XGW
acknowledges the support from A.P. Sloan Foundation.

\vfil\break

\end{document}